\documentclass[%
 reprint,
 superscriptaddress,
 amsmath,amssymb,
 aps,
]{revtex4-1}

\usepackage{graphicx}
\usepackage{dcolumn}
\usepackage{bm}

\begin{document}

\preprint{APS/123-QED}

\title{
	Three dimensional spectrometer
	}

\author{Muhammad Firmansyah Kasim}
\affiliation{
 John Adams Institute, Denys Wilkinson Building, Keble Road, Oxford OX1 3RH, United Kingdom
}
\author{Peter Norreys}
\affiliation{
 Clarendon Laboratory, Department of Physics, University of Oxford, Oxford OX1 3PU, United Kingdom
}
\affiliation{
 STFC Rutherford Appleton Laboratory, Chilton, Didcot OX11 0QX, United Kingdom
}
\date{\today}

\begin{abstract}
We present a novel design of 3D spectrometer that can retrieve 3D spectral profile in a single measurement. The 3D spectrometer design is built upon the concept of compressed sensing to make it possible to retrieve 3D information from 2D data from a screen/camera. In contrast to common spectrometers, the 3D spectrometer uses a wide slit instead of a narrow slit and retrieve the 3D datacube that consists of 2D spatial and 1D spectral information. Numerical tests were performed to simulate the retrieval of spectral profiles. The results show that the retrieved profiles match well the original profiles. It is also shown that the retrieved signal from the 3D spectrometer is robust enough for a further post-processing analysis.

\end{abstract}

\pacs{Valid PACS appear here}
\maketitle


\section{Introduction}
The development of compressed sensing \cite{comp-sense-candes, comp-sense-donoho, comp-sense-3, bayesian-comp-sense} in the recent decade has opened up new kinds of diagnostics and measurements in the field of science and engineering. A famous example is the single pixel camera \cite{single-pixel-camera} that would be useful for imaging in certain wavelengths where the detector array is expensive. Another compressed sensing application example in science is the 100 billion frame-per-second camera \cite{100-billion-fps} that was able to capture videos of light propagations.

Spectrometers are widely used in scientific community such as in the ultra short laser pulse \cite{spider} and laser-matter interaction fields \cite{sssi, fdh, fdsc, fdt, kasim-2}. The most common spectrometers used in the experiments with lasers are the Czerny-Turner spectrometer \cite{czerny-turner-spec} with narrow slits. With a narrow slit and 2D CCD camera, one can obtain the spectral profile of a slice of the light coming through the slit. This can be a problem when one wants to obtain the spectral profile of the light at the different slice, as they would need to do multiple measurements at multiple relative slit positions. Furthermore, there is no guarantee that they will get the same conditions for every measurements due to several factors, e.g. laser shot-to-shot variation. Sometimes these problems prevent experimentalists to do the multiple measurements.

By applying the concept of compressed sensing in the spectrometer, it is possible to retrieve multiple spectral profiles at different slices in a single shot. We call this as the 3D spectrometer.

\section{Compressed Sensing}\label{sec:compressed-sensing}
\subsection{Theory}
Consider a linear measurement system represented by the equation $\mathbf{y} = \mathbf{A} \mathbf{x}$, where $\mathbf{x}$ is an $N \times 1$ vector of the system's parameters to be measured, $\mathbf{y}$ represents an $M \times 1$ vector of measured data, and $\mathbf{A}$ is the measurement matrix of size $M \times N$. Retrieving $\mathbf{x}$ from the known measured data, $\mathbf{y}$, and the measurement matrix, $\mathbf{A}$, is known as the inverse problem.

If one has fewer measured data than the information to be retrieved (i.e. $M < N$), the problem becomes impossible to solve as there are infinite possible solutions. Some restrictions must be imposed in this under-determined system in order to get a reasonable solution.

In 2006, Candes, et al. \cite{comp-sense-candes} and Donoho \cite{comp-sense-donoho} stated that if $\mathbf{x}$ is a sparse vector, then it is possible to get the exact solution of an under-determined system. A vector is called sparse if it has only $k$ non-zero elements, where $k \ll N$. Even though the vector $\mathbf{x}$ is sparse, not all measurement matrices can be utilised to retrieve the vector $\mathbf{x}$. A good measurement matrix that can be used to retrieve the sparse vector $\mathbf{x}$ should have low coherence, where the coherence is defined as
\begin{equation}\label{eq:cs-coherence}
\mu(\mathbf{A}) = \max_{i\neq j, 1\leq i,j \leq N} \left( \frac{a_i^T a_j}{\|a_i\|_2 \|a_j\|_2} \right),
\end{equation}
with $a_i$ is the $i$-th column of matrix $\mathbf{A}$.

One key to achieve the low coherence in the measurement matrix is randomisation. It has been shown that a random Gaussian matrix will have low coherence with high probability \cite{comp-sense-donoho}. More recently, it has been shown that random Walsh-Hadamard matrix would work in practice as well \cite{single-pixel-camera}.

Sparsity is a key in compressed sensing. Even though, the signal of interest is not sparse in its original representation, most of the cases it would be sparse in some domain, e.g. DCT \cite{dct} or wavelet \cite{wavelet}. If this is the case, the measurement process can be expressed as $\mathbf{y} = \mathbf{A\Phi c}$, where $\mathbf{c}$ is the coefficient vector of signal $\mathbf{x}$ in other domain and $\mathbf{\Phi}$ is the sparsifying transformation matrix from the sparse domain to the signal's original representation. The matrix $(\mathbf{A\Phi})$ is treated as the measurement matrix in this case.

Assuming the coefficient vector, $\mathbf{c}$, is sparse, one way to retrieve $\mathbf{c}$ is by minimising the loss function below \cite{twist, fista, owlqn},
\begin{equation}
\label{eq:optimisation-problem-l1}
\mathcal{L} = \frac{1}{2} || \mathbf{y} - \mathbf{A\Phi c} ||_2^2 + \lambda || \mathbf{c} ||_1,
\end{equation}
where $\lambda > 0$ is the regulariser coefficient and $|| \mathbf{w} ||_p = \left(\sum_i |w_i|^p\right)^{1/p}$ is the $L_p$-norm of a vector. Another way is to retrieve the interesting signal directly, $\mathbf{x}$, using the total variation (TV) \cite{twist} as the regulariser instead of $L_1$-norm,
\begin{equation}
\label{eq:optimisation-problem-tv}
\mathcal{L} = \frac{1}{2} || \mathbf{y} - \mathbf{A\Phi c} ||_2^2 + \lambda \mathrm{TV}(\mathbf{x}),
\end{equation}
where TV can be the isotropic and non-isotropic total variation. The isotropic and non-isotropic TV are respectively given by \cite{twist},
\begin{align}
\mathrm{TV_{iso}}(\mathbf{x}) &= \sum_i \left[(\Delta_h x_i)^2 + (\Delta_v x_i)^2 \right]^{1/2}
\\
\mathrm{TV_{niso}}(\mathbf{x}) &= \sum_i \left( |\Delta_h x_i| + |\Delta_v x_i| \right),
\end{align}
with $\Delta_h$ and $\Delta_v$ respectively denote the gradient in horizontal and vertical directions.

Various algorithms are already available in solving the optimisation problems \ref{eq:optimisation-problem-l1} and \ref{eq:optimisation-problem-tv}. Among them are Two Steps Iterative Shrinkage Thresholding algorithm (TwIST) \cite{twist}, Fast Iterative Shrinkage Thresholding Algorithm (FISTA) \cite{fista}, Orthant-Wise Limited-memory Quasi-Newton (OWL-QN) \cite{owlqn}, and Iterative Hard Thresholding (IHT) \cite{iht}. They are gradient-descent based algorithms where a thresholding is applied after each descending step.

\subsection{Related works}

The most related works are perhaps Coded Aperture Snapshot Spectral Imaging (CASSI) with single disperser (SD-CASSI) \cite{sd-cassi} and double dispersers (DD-CASSI) \cite{dd-cassi}. Both works used disperser elements and a coded aperture to retrieve 2 dimensional spatial and 1 dimensional spectral profile of a scene. In DD-CASSI, the coded aperture is placed between two disperser elements. A disperser shears the light's spectral profile, then the coded aperture blocks some portion of the light in spatial and spectral domain, and finally the other disperser shears back the light's spectral profile. A 2D detector array is employed to detect the incoming light. From a 2D image captured by the detector, the 2D spatial and 1D spectral profile of the light can be retrieved.

SD-CASSI has the similar configuration to DD-CASSI, except that it does not have the first disperser element. Some portion of light coming from a scene will be blocked, then be sheared by the disperser, and finally be recorded on the detector. Both configurations, SD-CASSI and DD-CASSI, are similar, but SD-CASSI has the advantage of getting higher spectral resolution while DD-CASSI can get advantage if the spatial resolution is more demanded than the spectral resolution.

In SD-CASSI, a measurement is done in spectral and spatial domain, while the randomisation is applied to spatial domain only. That makes the measurement matrix in SD-CASSI has reasonably high coherence for measurements that demand much more spectral information than spatial information. Therefore, for some spectrometer applications, such as interferometry, which require high accuracy and high resolution in spectral profile as well as a post-processing analysis, SD-CASSI might give undesirable results.

Our work introduces a spectrometer design that gives high accuracy and resolution of the spectral profile, better than SD-CASSI, and it is also robust for applications where a post-processing of the signal is required.

\section{3D Spectrometer}\label{sec:3d-spectrometer}
\begin{figure*}
\includegraphics[scale=0.23]{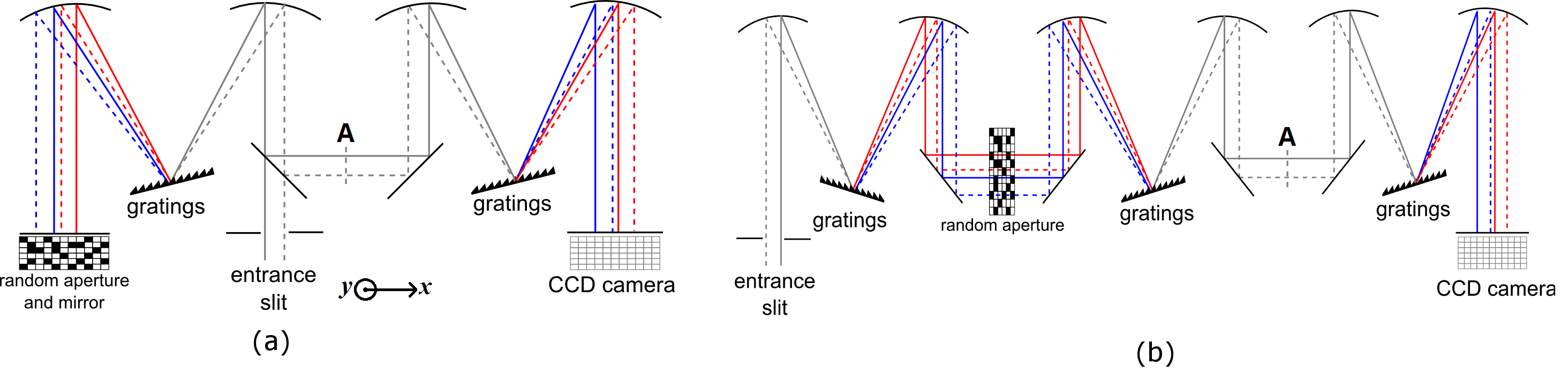}
\caption{\label{fig:schematics} Schematic design of the 3D spectrometer. Solid lines show the path of the ray entering from one side of the slit while the dashed lines for the ray on the other side. Blue and red lines are respectively representing high and low frequency components of the light. The $x$-axis is horizontal and the $y$-axis is coming out of the plane. The left picture (a) is the folded schematics of the design and the right picture (b) is the unfolded schematics.}
\end{figure*}

In order to achieve 2D spatial and 1D spectral measurement with high spectral accuracy, we propose a novel spectrometer design. The schematic of this spectrometer is shown in Fig. \ref{fig:schematics}(a).

The design consists of 2 Czerny-Turner spectrometers \cite{czerny-turner-spec} with a beam splitter to connect these 2 parts. On one side, there is a random aperture with a mirror to block some part of the light and reflect it back. This can be replaced by the Digital Micromirror Device (DMD). And on another side, there is a detector array to record the intensity of the light on that plane. One can use Charge-Coupled Device (CCD) camera as the detector. Another difference of this design with other spectrometers is that the 3D spectrometer uses a wide slit instead of a narrow slit.

\begin{figure}
\includegraphics[scale=1.0]{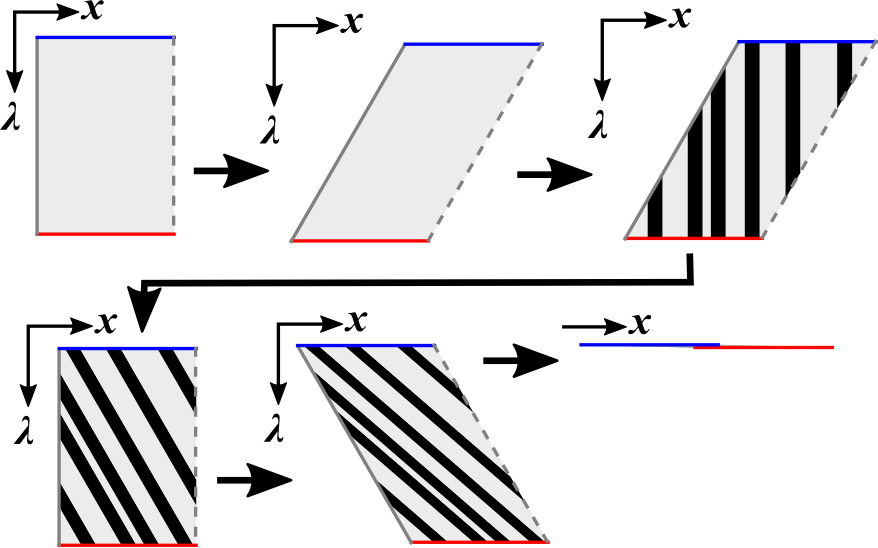}
\caption{\label{fig:datacube-transformation} The datacube transformation of the light from entering the slit until reaching the CCD camera. The $x$-axis is corresponding to the axis in Figure \ref{fig:schematics}(a) and the $y$-axis is coming out of the plane. The $\lambda$ axis on the datacube represents the spectral axis of the light's profile.}
\end{figure}

To retrieve the 3D information (2D spatial and 1D spectral profile) of the light entering the slit, it is essential to know how the 3D profile of the light is transformed in the spectrometer, or the datacube transformation. The datacube transformation of the light in this spectrometer design is shown in Figure \ref{fig:datacube-transformation}.

From the entrance slit, the datacube is sheared by the gratings on the left side of the picture. After the light is reflected back by the random aperture and mirror, some part of the datacube is blocked and sheared back by the same gratings when it reaches plane A in Figure \ref{fig:schematics}. Then by another gratings, the datacube is sheared to the opposite direction before it reaches the detector array. Because the detector array can only record the total intensity as function of positions, the datacube is then summed over the $\lambda$-axis. From this recorded intensity profile on the detector, one can employ a compressed sensing retrieval algorithm to obtain the original datacube, such as TwIST \cite{twist}, FISTA \cite{fista}, OWL-QN \cite{owlqn}, and IHT \cite{iht}.

In this spectrometer design, the detector measures the intensity in spectral and spatial domain, while the randomisation also takes place in spatial and spectral domain. This gives more randomisation in the measurement, and thus it has high probability to get lower coherence and better retrieval ability compared to the previous methods \cite{sd-cassi, dd-cassi}.

An equivalent design of this schematic is shown in Figure \ref{fig:schematics}(b) where the design is unfolded. The unfolded design gives an advantage of higher light flux reaching the CCD camera. In the folded design (Figure \ref{fig:schematics}(a)), at least 25\% of the light is loss in the beam splitter. This does not happen in the unfolded design because it does not involve a beam splitter. However, more gratings and mirrors are required for this unfolded design.

\section{Numerical tests}\label{sec:results}
In order to test the design, we performed some numerical tests. The tests were performed in Python and the signals were retrieved using ANOA library in Python. It is a differentiable programming library we wrote that automatically provides the gradients of a variable with respect to some variables. It also includes some compressed sensing algorithms such as FISTA, TwIST, and OWL-QN. In this case, the program determines the gradients of the loss from equation \ref{eq:optimisation-problem-l1}, $\mathcal{L}$, with respect to the spectral cube's elements, $\mathbf{c}$. The program also determines $\mathbf{c}$ that minimises the loss $\mathcal{L}$.

A code is written to simulate the datacube transformation using concave toroidal mirrors with focal lengths of $f = 15\ \mathrm{cm}$ and gratings with 300 lines/mm. The random aperture in the simulation has $100 \times 100$ square pixels with side length of $20\ \mathrm{\mu m}$ each. The detector array has $20\ \mathrm{\mu m}$-size square pixels with $100 \times 100$ pixels.

The first case uses a spatially chirped light source, where the light has different central frequency for different $x$-position in this case. For every $x$-position, the light has a Gaussian spectral profile with the full width half maximum (FWHM) is 4.05 nm and the central wavelength is around 633 nm. The chirp is 31.25 nm/mm. Gaussian noise with $\sigma$ 10\% of the maximum spectral intensity is also added to test the robustness. Width of the slit is $200\ \mathrm{\mu m}$ or about 10 times of the detector's pixel size.

For the retrieval algorithm, OWL-QN algorithm is employed \cite{owlqn} with the datacube size of $(N_y, N_x, N_\lambda) = (100 \times 11 \times 90)$ voxels. OWL-QN is a second order optimisation algorithm for $L_1$ regulariser. Three dimensional Discrete Cosine Transformation (3D-DCT) domain is chosen as the signal's sparse domain.

\begin{figure*}
\begin{center}
\includegraphics[scale=0.37]{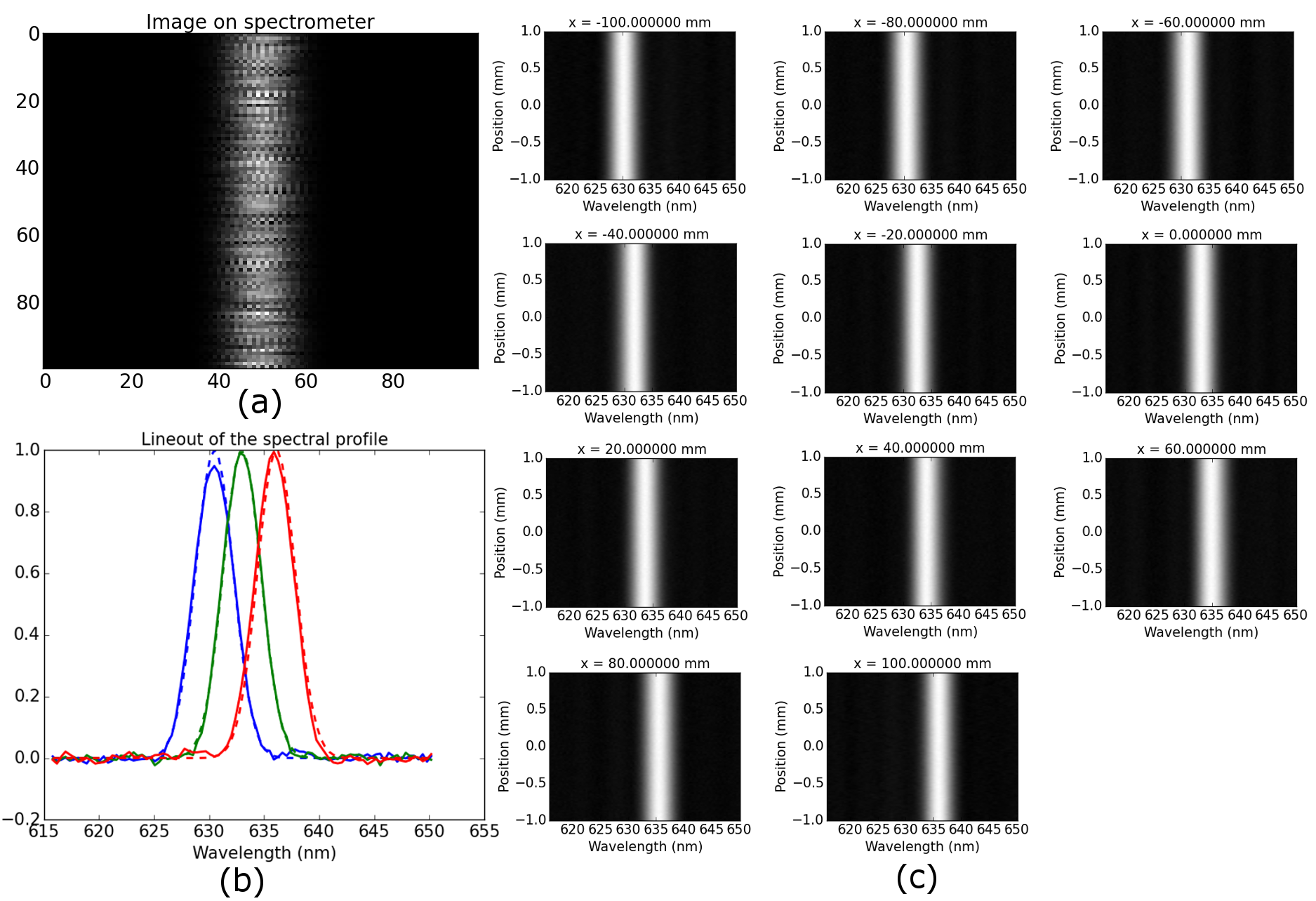}
\caption{\label{fig:chirp-position-all} (a) The image captured on the CCD camera in the 3D spectrometer. (b) Lineout of the spectrum from the leftmost slice (blue), middle slice (green), and the rightmost slice (red). (c) The spectral intensity profiles for every slices.}
\end{center}
\end{figure*}

From the measured intensity profile shown in Figure \ref{fig:chirp-position-all}(a), the complete 3D datacube profile of the light entering the slit can be retrieved, as shown in Figure \ref{fig:chirp-position-all}(c). The figures clearly show the central wavelength of the light is different for slices in different $x$-position. As shown in Figure \ref{fig:chirp-position-all}(b), the central lineouts of the retrieved signal agree very well with the original signal. The spectral profiles in Figure \ref{fig:chirp-position-all}(c) are also a great agreement with the original profiles.

Another test was done based on spectral interferometry case \cite{sssi, fdh, kasim-2}. In the case of spectral interferometry, there are two short laser pulses. One pulse, called as the probe pulse, has some phase modulation while another pulse, the reference pulse, does not have phase modulation. The pulses are temporally chirped. They are co-propagating with each other and temporally separated. The spectral profile of these two pulses contains information of the phase modulation of the probe pulse. By applying a high pass filter and obtaining the phase of the spectral profile, one can extract the phase modulation of the probe pulse. Implementing this in the 3D spectrometer poses a greater challenge since the spectral interferometry needs a post-processing (i.e. phase extraction) of the retrieved signal to obtain useful information.

In this simulated case, the phase of the probe pulse is modulated spatially and spectrally. 
The modulation is then encoded into the spectral intensity according to the equation below,
\begin{equation}
I(x,y,\omega) = I_0(x,y,\omega) \left[1 + \cos\left(\omega\tau + \Delta \tilde{\phi}(x,y,\omega)\right)\right],
\end{equation}
where $I_0$ is the spectral intensity profile for one pulse and $\tilde{\phi}(x,y,\omega)$ is the phase modulation of the probe pulse. In the 3D spectrometer, the spectral intensity datacube, $I(x,y,\omega)$, undergoes the transformation as in Figure \ref{fig:datacube-transformation} with $\lambda = 2\pi c/\omega$. The spectral intensity datacube is retrieved from the image captured by the 3D spectrometer and further the post-processing analysis is performed to obtain the phase modulation datacube, $\tilde{\phi}(x, y, \omega)$.

\begin{figure}
\begin{center}
\includegraphics[scale=0.25]{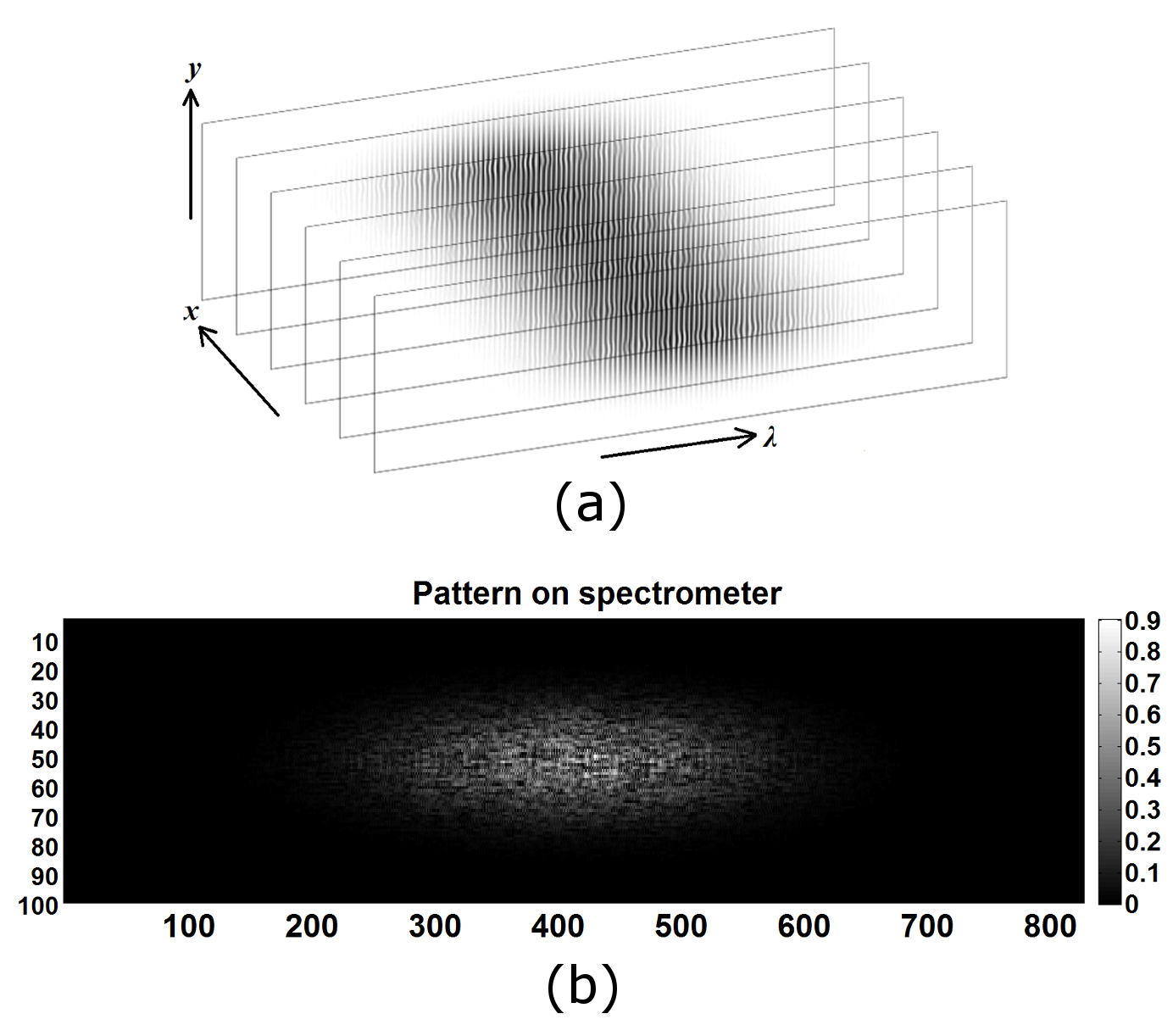}
\caption{\label{fig:interferometry-spectrometer} (a) Illustration of the datacube to be inverted. There are only $N_x=6$ slices shown here, but in the test case, $N_x=10$ is used. (b) The simulated pattern captured on the screen in the 3D spectrometer.}
\end{center}
\end{figure}

\begin{figure*}
\begin{center}
\includegraphics[scale=0.25]{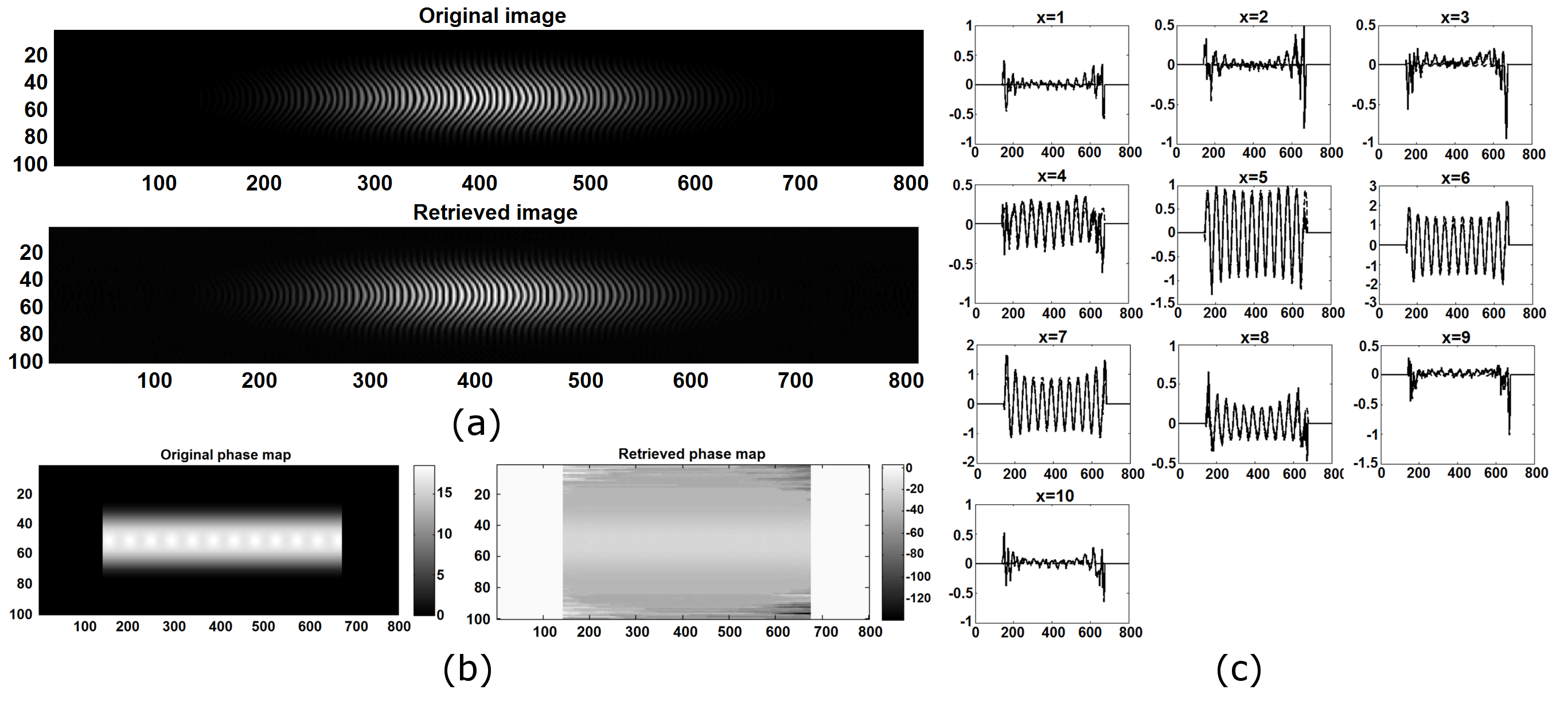}
\caption{\label{fig:interferometry-results} (a) Comparison of a slice of original spectral intensity profile and the profile retrieved from \ref{fig:interferometry-spectrometer}(b). There are 10 slices of the retrieved profiles in total. (b) The comparison between the extracted phase map from the original and the retrieved spectral profiles. (c) The central lineout of the phase map from every slice. Dashed lines are the original phase map profile and the solid lines are the retrieved phase map. Both profiles are in a great agreement.}
\end{center}
\end{figure*}

In the tested case, there are $N_x=10$ slice of the spectral intensity profiles to be retrieved each with size $N_{\lambda} \times N_y = (800\times 100)$ pixels. The datacube example is shown in Figure \ref{fig:interferometry-spectrometer}(a). The pattern on the 3D spectrometer's screen is shown in Figure \ref{fig:interferometry-spectrometer}(b). The size of the image on the screen is $(809 \times 100)$ pixels. Retrieving all the spectral intensity profiles was done using TwIST algorithm. One of the retrieved spectral intensity profiles and its comparison with the original spectral intensity profile is shown in Figure \ref{fig:interferometry-results}(a).

One advantage of this 3D spectrometer design is that it is also robust even with a post-processing. Figure \ref{fig:interferometry-results}(b) show the extracted phase maps from the retrieved and original spectral profiles. The comparison for the central lineouts of the phase maps from all the spectral intensity profiles is shown in Figure \ref{fig:interferometry-results}(c). This shows an excellent agreement between the retrieved phase maps and the original phase maps.

\section{Conclusions}\label{sec:conclusions}
A robust design of 3D spectrometer is presented. In the 3D spectrometer, a wide slit is used instead of a narrow slit. The 3D spectrometer can retrieve the spectral profiles of the light coming through the slit. It is shown that the design is robust even with noise and also robust for any signal that requires a post-processing.

This opens up new possibilities of diagnostics such as capturing a high speed video ($\sim$trillion fps) for laser-matter interaction, complete spectral profile analysis of a laser pulse after the interaction with matter, and if combined with SPIDER \cite{spider}, the 3D phase and intensity profiles reconstruction of a short pulse is possible.

\end{document}